# Ultra-compact optical auto-correlator based on slow-light enhanced third harmonic generation in a silicon photonic crystal waveguide

**Christelle Monat*[1,2], Christian Grillet[1,2], Matthew Collins[2], Alex Clark[2], Jochen Schroeder[2], Chunle Xiong[2], Juntao Li[3], Liam O'Faolain[3], Thomas F. Krauss[3**], Benjamin J. Eggleton[2], and David J. Moss[2***]**

[1] *Université de Lyon, Institut des Nanotechnologies de Lyon (INL), Ecole Centrale de Lyon, 69131 Ecully, France*
[2] *CUDOS, Institute of Photonics and Optical Science (IPOS), School of Physics, University of Sydney, New South Wales 2006, Australia*
[3] *School of Physics and Astronomy, University of St Andrews, St Andrews, Fife, KY16 9SS, UK*

** *T.F. Krauss current address: Department of Physics, University of York, York YO10 5DD, UK.*
*** *D.J. Moss current address: School of Electrical and Computer Engineering, RMIT University, Melbourne, Vic., Australia 3001*
*christelle.monat@ec-lyon.fr

**The ability to use coherent light for material science and applications is directly linked to our ability to measure short optical pulses. While free-space optical methods are well-established, achieving this on a chip would offer the greatest benefit in footprint, performance, flexibility and cost, and allow the integration with complementary signal processing devices. A key goal is to achieve operation at sub-Watt peak power levels and on sub-picosecond timescales. Previous integrated demonstrations require either a temporally synchronized reference pulse, an off-chip spectrometer, or long tunable delay lines. We report the first device capable of achieving single-shot time-domain measurements of near-infrared picosecond pulses based on an ultra-compact integrated CMOS compatible device, with the potential to be fully integrated without any external instrumentation. It relies on optical third-harmonic generation in a slow-light silicon waveguide. Our method can also serve as a powerful in-situ diagnostic tool to directly map, at visible wavelengths, the propagation dynamics of near-infrared pulses in photonic crystals.**

In the last decade, silicon photonics has enabled the development of a wide range of integrated all-optical signal processing devices [1] on CMOS compatible chips that have achieved an impressive array of new capabilities, from switching and de-multiplexing at unprecedented speeds [2], to parametric gain [3], Raman lasing [4], optical logic [5], regeneration [6], terahertz bandwidth radio frequency (RF) spectrometers [7], to temporal optical cloaking [8], and many others. However, an integrated chip-based pulse measurement device that is capable of single-shot measurements, without the need for a reference pulse, averaging or some external instrumentation that is difficult to integrate, such as an optical spectrum analyser or a long tunable delay line, is still lacking. While several groups have replicated all-optical techniques on a chip [9-12] that are well established in free-space or fiber-optics, such as autocorrelation, frequency-resolved optical gating (FROG), or temporal sampling, with obvious benefits in footprint, energy consumption and performance, all of the demonstrations fall short in one of the aforementioned points. A fully integrated pulse-measurement technology is nonetheless highly attractive, enabling the complete integration of pulse-generation, optical processing and measurement schemes on a single chip. This would provide an extremely flexible integrated processing platform, in particular in conjunction with integrated pulse-shaping [13] or arbitrary waveform generation [14].

The silicon-on-insulator (SOI) technology offers many attractive features for nonlinear optics [1]. Its high refractive index allows tight confinement of light within nanowires that, combined with the high silicon Kerr nonlinearity ($n_2$), has yielded extremely high nonlinear parameters ($\gamma$) of ~ 300W$^{-1}$ m$^{-1}$, i.e. 4 to 5 orders of magnitude larger than in silica fibers ($\gamma = \omega\, n_2 / c\, A_{eff}$, where $A_{eff}$ is the effective area of the waveguide, $c$ is the speed of light and $\omega$ is the optical signal frequency). Furthermore, SOI has proven to be a powerful platform for enhancing nonlinear optical effects over short propagation distances (well below 1mm) by increasing the optical beam intensity via so-called "slow light" in planar photonic crystal (PhC) waveguides [15,16]. In contrast to resonant cavities that only accommodate very limited optical bandwidth, advanced slow light PhC waveguide designs have been able to offer both wide-bandwidth and relatively low dispersion, i.e. "flatband" operation [17] with relatively low propagation losses [18]. This is critical for processing short optical signals since slow light is otherwise associated with the dispersive band-edge of PhC waveguides which tends to strongly broaden and distort short optical pulses [19]. By enhancing the effective Kerr nonlinearity in this manner, slow light planar

PhC waveguides have yielded much more compact devices [20-24] that are compatible with high speed signals [24,25]. In particular, slow light enhanced optical third harmonic generation (THG) was reported in silicon PhC waveguides [22] and subsequently exploited to monitor the quality of 640Gb/s optical signals [24].

Here, we report a novel application of THG in slow light silicon PhC waveguides to achieve single shot temporal measurements via the auto-correlation of picosecond optical pulses. The visible light signal emitted off the chip effectively maps the temporal characteristics of the near-infrared pulse into spatial dimensions which can then easily be acquired with an out-of-plane camera or detector array. A number of integrated all-optical techniques have already been reported [9-12,26,27] to achieve time-resolved measurements of ultra-high speed optical signals as an alternative to expensive, bulk-optics based, commercial optical auto-correlators. These include time-lens temporal imaging on a silicon chip [9, 26], and for phase-sensitive pulse measurements, waveguide based frequency resolved optical gating (FROG) [10,27] and a novel variation of spectral phase interferometry for direct electric-field reconstruction (SPIDER) [11]. In contrast to these techniques, our demonstration does not rely on off-chip optical spectrum analysers [9, 11,26], nor a complicated numerical extraction algorithm [11]. Further, no reference pulse or long tunable delay lines [10,12] are required, and unlike sampling approaches [10], there is no need to average over many pulses. The simple and extremely compact (96μm long) device reported here opens the door to the realization of the first fully integrated all-optical device capable of single shot high-speed optical signal diagnostics in the near-infrared telecommunications window. The technique also produces the equivalent of time-of-flight measurements of picosecond pulses as they propagate through slow light photonic crystal waveguides. This capability provides a powerful in-situ diagnostic tool that allows us to obtain new insights into the propagation dynamics of a slow light pulse and to directly extract its group velocity.

## Experiments

**Device Principle.** The principle of operation for the auto-correlator and the experimental set-up for characterizing the device performance are shown in Figure 1. When near-infrared pulses (~1550nm) propagate along the waveguide independently from either side, they produce THG, i.e. visible light at the third-harmonic (TH) wavelength (~530nm), due to the ultrafast $3^{rd}$ order optical nonlinearity in the silicon PhC waveguide [22]. The efficiency of this process is significantly enhanced by the large (typically ~ 30) group index in these slow light waveguides [22]. The spatial profile of THG-induced visible light along the waveguide is relatively uniform for each pulse separately, but when the counter-propagating pulse replicas are made to overlap within the PhC waveguide, an additional THG component -referred to as cross-THG- produces a spatially varying intensity profile along the slow light PhC waveguide. This stationary pattern of visible light is proportional to the interference of the two pulse replicas – i.e., it effectively maps the temporal autocorrelation of the pulse into space, in a way similar to the free-space optics scheme demonstrated in a SHG based bulk crystal with kilowatt peak power pulses [28]. As represented on Fig. 1(b), imaging this spatial intensity profile then yields the temporal autocorrelation of the pulse, potentially for a single pulse, or so-called "single shot" operation – a key advantage over other approaches that are sampling based, such as FROG, and rely on averaging over thousands or even millions of pulses.

Figure 1(a) shows the experimental configuration for characterising the device. Optical pulses ~2.5ps wide from a tunable mode-locked fiber laser are split in two and launched into opposite ends of the PhC waveguide (see Methods). The relative delay between the two counter-propagating pulses can be adjusted in order to control whether or not the pulses overlap in time while propagating in the PhC waveguide. A high N.A. (0.65) objective and a CCD visible wavelength camera are used to collect and image the TH signal emitted along the PhC waveguide from the top of the chip. The spatial profile of the TH signal, imaged onto the CCD detector, thus directly converts the temporal profile of the pulse autocorrelation into a spatial intensity profile. Note that while the two arms of the set-up include numerous off-chip components to control polarization, circulators and attenuators to balance the optical power of the two beams launched from either side, as well as a tunable delay line in one arm to match the optical path lengths, most if not all of these components are employed just to provide a clear proof of principle of the device operation. They could all readily be avoided with the simple use of an integrated 3dB splitter on-chip as was successfully demonstrated for SWIFT integrated spectrometers (Stationary-wave integrated Fourier-transform spectrometry) [29].

**Device architecture.** Figure 2 shows a schematic of the slow light PhC waveguide along with an SEM picture of the device and the group index versus wavelength measured by interferometric techniques [30]. The device consists of a 96 µm long silicon PhC waveguide in a 220 nm-thick membrane suspended in air (see Methods). It is connected on both sides to silicon ridge waveguides terminated by inverse tapers and embedded in polymer waveguides to provide a total insertion loss of ~ -9.5 dB (fiber to fiber). The PhC waveguide is created from a triangular lattice of air holes (period ~404 nm, radius~116 nm), with one row omitted along the ΓK direction and the first surrounding rows laterally shifted to engineer the waveguide dispersion [17]. Unlike the highly dispersive slow light mode associated with the band edge of typical PhC waveguides [19], here the fundamental mode is engineered to display both low group velocity *and* low dispersion. As displayed on Fig. 2(b), the resulting group velocity in this device is ~c/30 within 10% over a bandwidth exceeding 15 nm, i.e. readily compatible with the use of high-speed Tb/s signals [24].

## Results and Discussion

Figure 3 shows a series of images recorded on the CCD camera when pulses at 1548 nm are launched into the waveguide from one side only (Fig. 3(b) and (c)) and from both sides with the delay line adjusted so that the two pulses overlap either off chip (Fig. 3(d)) or in the middle of the PhC waveguide (Fig. 3(e)). For cases where the pulses do not overlap on the chip, the TH green light signal is relatively uniform with a low power level, which effectively provides a relatively weak and uniform autocorrelation background. By contrast, when the delay is adjusted such that the pulses overlap inside the PhC waveguide, the average power of TH light is much larger and the associated spatial intensity profile exhibits a stationary Gaussian-like shape centered around the middle of the PhC waveguide. This pattern results from cross-THG produced from the interaction between the left and right near-infrared pulses in the PhC waveguide. From the spatial profile of the TH-power extracted from the CCD image, we retrieve the temporal pulse intensity profile, taking into account the ~c/23 group velocity at which the 1548nm wavelength pulses propagate in the PhC waveguide (see Methods). This intensity profile is compared on Fig. 3(f) with that measured at the output of the laser using a commercial second-harmonic generation based auto-correlator. On this time-resolved plot, the physical boundaries of the slow light PhC waveguide coincide with the points at around +/-6ps, outside of which no TH-light is recorded but just noise. The full-width at half maximum (FWHM) inferred from the THG profile is ~2.4 ps, agreeing relatively well with 2.6 ps obtained via the commercial auto-correlator. The deviation at both edges of the profile is due to the residual TH average signal generated when the two pulses are launched into the waveguide separately from either side. Although it is lower than for SHG based autocorrelation [31], this background does reduce the signal to noise ratio. In our case though, it can be completely suppressed by using appropriate spatial filtering.

Figure 4 illustrates the possibility to implement such a spatial filtering scheme by imaging the TH output in the Fourier plane [32,33]. Additional cross-terms in k-space (extreme side lines on Fig. 4(b)) are observed when the optical pulses overlap in the PhC waveguide. These are due to the interaction of two left (respectively right) fundamental photons with one right (respectively left) photon to generate one TH photon, instead of classical THG where all fundamental photons come from the same side. The momentum conservation of the photons involved in the nonlinear process are represented on Fig. 4(a) and (b) for THG and cross-THG, explaining the different lines observed on Fig. 4 (at $\pm (3k_\omega - 2\pi/a)$ for THG and $\pm k_\omega$ for cross-THG). Note that unlike free-space schemes that typically require accurate angle tuning, this quasi-phase matching for cross-THG is naturally ensured by the planar PhC waveguide geometry, independently of the frequency of the fundamental beam [32]. In principle, spatially filtering out the k components closer to the optical axis would eliminate THG induced by the separate pulses and yield a background-free autocorrelation trace. As expected, the cross-THG signal observed in the Fourier plane on Fig. 4(b) is more intense that the THG. Indeed, the overall intensity of the combined left and right pulses is larger than their separate intensities, while the THG process scales with the cube of the intensity at the fundamental frequency. Note, however, that for our specific waveguide geometry, the fundamental mode wavevector ($k_\omega \approx 0.3 \, [2\pi/a]$) positions the cross-THG lines ($k_{3\omega} = \pm k_\omega$) off-center of the objective N.A. (represented by the dashed white circle on Fig. 4). This decreases the collection of the relevant cross-THG relative to the separate THG process, effectively increasing the THG background superimposed on the auto-correlation trace observed in the real TH images.

In addition to providing the nonlinear enhancement sustaining the THG process, slow light propagation spatially compresses the pulse [34], providing the basis for more compact devices. Using the device in regions having different group velocities also enables the measurement of a wide range of pulse durations within a very short waveguide by effectively adjusting the sensitivity and temporal resolution of the device. We investigate the effect of group velocity on the device performance, by studying the behavior of the device in a regime where the group velocity of the fundamental mode varies almost linearly, by a factor of 4, from c/8 to c/30 between 1540 nm and 1550 nm (see Fig. 2(b)). Figure 5(a) shows the measured TH spatial profiles obtained when launching pulses with similar durations tuned to slightly different wavelengths. We fit the TH patterns with Gaussian curves (see Fig. 5(b)) and plot the associated full-width at half maximum (FWHM) on Fig. 5(c), which follows the expected linear trend with the inverse of the group index [34]. We thus directly observe the spatial pulse compression effect for decreasing group velocities. Selecting the waveguide group velocity can therefore be used to tune the auto-correlator performance, through adjustment of the device calibration, temporal resolution and sensitivity (see Table 1 in the Methods section). For instance, with the present ~100μm long slow light waveguide, pulses with a maximum duration of 17ps can be readily measured with a resolution greater than ~69fs, while a resolution better than 20fs is available at a lower (c/8) group index. On the short range side, pulses as short as 550fs can be measured, limited here by the dispersion in the flat-band slow light window of the waveguide under test.

A novel and significant feature of our approach is that, conversely, it can serve as a diagnostic tool to directly measure and study the near-infrared pulse propagation and group velocity in the slow light regime of a PhC waveguide - i.e. in a regime where measurements become difficult. This can be achieved by sweeping the off-chip delay line and tracking the peak of the TH profile along the waveguide. Figure 6(a) shows the equivalent of time-of-flight measurements, with a series of snapshots of the spatial (thus temporal) profile of the pulse as it "propagates" across the chip. The series of images is obtained by varying the relative temporal delay between the two incident pulses in the right and left arm, within 10ps, which effectively changes the location at which the two pulses meet within the PhC waveguide. From this, we can extract the group index for different wavelengths, in a way that is much more direct and easier than traditional interferometric measurements, which typically require high stability and data post-treatment. Figure 6(b) shows the spatial shift of the TH profile versus temporal delay at two different wavelengths (1545 nm and 1552 nm), corresponding to two different group velocities. Since the TH peak location corresponds to the point at which the two pulses overlap within the PhC waveguide, the group index can be inferred, without any fit parameter or assumption, from the slope of these linear curves, resulting in an $n_g$=15 and 28 at the two probed wavelengths (see Methods). Figure 6(c) presents the mapped group index measured at 6 different wavelengths, overlaid with the group velocity dispersion obtained by interferometric means [30], showing good agreement between the two. The measurement of the group velocity and the slow light induced spatial compression of the pulse are just two examples of in-situ information related to the pulse propagation that can be retrieved. Additional key information relates to the capability of the pulse to preserve its integrity at low speeds, where the ballistic regime has been questioned in the presence of disorder [35,36]. This time-to-space mapping technique combined with a tunable delay line therefore enables the direct observation of the pulse propagation dynamics as an alternative to powerful but relatively slow and complex heterodyne NSOM techniques [34,37].

In the present demonstration, the key component - the waveguide based time-to-space mapping element - is the PhC waveguide which is ultra-compact (<100μm long) and integrated on chip. A fully integrated optical auto-correlator could be achieved by integrating a 3dB splitter as well as the visible imaging system on-chip, thereby replacing the external lens/ CCD camera. To accomplish this, our device could be integrated on top of a silicon based linear detector array that would collect the THG signal directly in the near-field of the PhC waveguide, without lens assistance. Since the generated TH light is in the visible, it is therefore detectable with standard silicon detectors. The successful integration of near-infrared SWIFT spectrometers with a complicated set of sensitive single photon detectors [29] points to the feasibility of this approach. This would yield a fully integrated device fabricated with solely CMOS technology, and without the need for any external instrumentation. The use of integrated detector arrays may also greatly improve the detection efficiency, allowing us to lower the required peak power level in the sub-Watt regime, potentially down to the 10's of milliwatt regime. A full analysis of the implications of this on the detector sensitivity and resulting temporal resolution, however, lies outside the scope of this paper. Finally, it should be mentioned that, unlike SPIDER for example, the present device does not recover the phase information of the pulse. While in many cases the auto-correlation pulse-shape is sufficient, the device performance could be further enhanced by exploiting on-chip pulse compression techniques [38] that provide a shorter reference pulse from the signal pulse itself in order to generate the spatial pulse intensity profile directly rather than the autocorrelation.

While the performance of our device is not quite that of commercial bulk optics based auto-correlator, (our temporal resolution is slightly larger (~20fs versus 5fs for FR-103XL by Femtochrome while our time window is slightly less (~17ps vs 90ps)), the compact nature and potential for full integration is a major advantage that will enable many new applications. We note also that our performance can readily be improved by adopting different designs. For example the 17ps window can be extended up to 340ps using a 2mm long integrated waveguide without degradation of the temporal resolution (see Methods).

While it may appear that a drawback of this device is its bandwidth, or wavelength range (15nm versus an octave for a commercial instrument), we first note that the bandwidth of this device (at well over 1THz) is well beyond that of other nonlinear devices that employ resonantly enhanced nonlinearities. Secondly, the extreme compactness of the slow light waveguide - more than two orders of magnitude shorter than a typical centimeter scale commercial auto-correlator – would allow the integration (with no moving parts) of parallel devices at staggered wavelengths in order to cover any wavelength range of interest. The C-band, for example, could be entirely covered by only 2 to 3 waveguides by splitting the signal across several spectrally detuned slow light PhC waveguides with the CCD camera monitoring the auto-correlation signals produced by adjacent waveguides simultaneously. The operation wavelength of the flat-band slow light window can be indeed finely adjusted during the lithography process [17]. Further, integrating a high resolution spectrometer such as an AWG (readily available with silicon photonics technology) with an array of PhC waveguides on chip would allow the application of this device to WDM signals of up to 50 or more channels.

In summary, we have demonstrated a THG-based optical autocorrelator that employs an integrated and ultra-compact slow light silicon PhC waveguide of less than 100μm in length. Picosecond pulses near 1550nm are imaged directly along the PhC waveguide, providing reliable and potentially single-shot measurements of the pulse. We also demonstrate that this technique can be used as a simple and effective in-situ diagnostic tool to study the propagation dynamics of near-infrared pulses into slow light photonic crystal waveguides. This demonstration opens the door to realizing fully integrated devices capable of measuring ultra-fast near-infrared pulses on a silicon chip.

# Methods

**Device structure and fabrication.** The device was fabricated from a SOITEC silicon-on-insulator wafer by electron-beam lithography (hybrid ZEISS GEMINI 1530/RAITH ELPHY) and reactive ion etching using a $CHF_3/SF_6$ gas mixture. The silica layer under the PhC slab was selectively under-etched using a HF solution to leave the PhC section in a suspended silicon membrane. The 2D PhC structure consists of a triangular lattice of air holes with lattice constant a=404 nm and hole radius 116 nm (0.287a) etched into a 220 nm thick silicon suspended membrane. A W1 linear waveguide is created by omitting a single row of holes along the ΓK direction. The total PhC waveguide length is 96 µm, and the lattice period of the first and last 10 periods is stretched by around 10% (to ~444 nm) parallel to the waveguide axis so as to enhance the coupling to the slow light mode. The dispersion of the PhC waveguide is engineered by laterally shifting the first rows of holes adjacent to the waveguide center in the direction perpendicular to the waveguide axis. For the dispersion engineered waveguide used in this experiment, the rows closest to the center are shifted 50 nm away from the axis of the waveguide and the mode effective area $A_{eff}$ is around 0.5 µm$^2$. The PhC waveguide is connected on both sides to 200 µm long access nanowires (220 nm × 700 nm cross-section) terminated by inverse tapers, and embedded into large mode area ($A_{eff}$~8 µm$^2$) SU8 polymer waveguides to improve the coupling to the chip via butt-coupled lensed fibres. The total fiber to fiber transmission is about -9.5dB in the present experiment. The absolute group velocity dispersion $\beta_2$ for the engineered fundamental PhC waveguide mode ranges between $3\times10^{-21}$ s$^2$/m, in the flat-band slow light window, up to $5\times10^{-20}$ s$^2$/m in the dispersive band between 1545 nm and 1550 nm. This provides an associated dispersion length for 2.5 ps pulses between 125µm and 2mm. This is longer than the entire PhC waveguide length, and so we expect the effect of dispersion on all our measurements to be negligible. In the flat-band slow light window with reduced dispersion, this value of $\beta_2$ enables, in principle, the measurement of pulses as short as 550 fs without distortion along the waveguide. Such short pulses could be readily mapped along the length of our device since they would have a spatial extension within the waveguide of ~5.5 µm, which is 20 times larger than the spatial resolution of our imaging set-up (see below).

**Optical Testing.** As shown in Figure 1, the experimental set-up consists of splitting optical pulses (2.5ps pulses at 10 MHz repetition rate, λ=1550 nm, ~10 W peak power) from a tunable mode-locked fiber laser and launching the resulting counter-propagating replicas onto the chip. The two arms of the set-up include polarisation controllers, attenuators and circulators to select the TE polarization and balance the optical power, measured onto power meters (through 1% couplers), arising from either side. We noticed that the TH pattern was not very sensitive to finding the perfect power balance though. One arm contains a motorized tunable delay line (with a maximum delay of 250 ps) to match the optical path lengths on both sides. Note that all of these components could be avoided with the use of an integrated 3 dB splitter. The average power launched on each side of the chip for the autocorrelation measurement presented on Fig. 3 (at 1548nm, $n_g$~23) is about 200µW, corresponding to around ~3.2W coupled peak power for -4dB estimated insertion loss. When tuning the signal wavelength as on Figs. 5 & 6 to explore the impact of the group velocity while maintaining a THG-based autocorrelation trace with a relatively high signal to noise ratio, the input power was slightly varied within 2W to 3.5W coupled peak power. This was done to compensate for (1) the decrease of the THG efficiency at lower group indices [22], (2) the slight increase in the propagation loss with the group index and (2) the decrease in the collection efficiency of the cross-THG signal for increasing wavevectors (i.e. larger wavelengths). A 0.65 N.A. Mitutoyo objective is used to collect the TH signal emitted from the top of the PhC waveguide and imaged onto a high resolution silicon CCD visible camera. A beam-splitter complements the free-space optic collection system to simultaneously image the waveguide in the Fourier plane on another visible CCD camera. The details on the Fourier space measurements are further explained in reference [32]. For the figures displaying the profile of the TH intensity, the TH signal is averaged over 17 pixels in the lateral direction centered on the waveguide axis.

**Time to space mapping and device resolution.** Our auto-correlator device effectively provides a time (T) to space (Z coordinate along the waveguide axis) mapping of the pulse temporal characteristics through the following relation: $T=Z \times c/n_g$ when accounting for the group velocity $c/n_g$ of the pulse into the PhC waveguide. In addition, the cross-THG intensity image provides a temporal profile that is compressed by a factor $\sqrt{3}$ with respect to the pulse intensity profile, as estimated from numerical simulations of the THG produced by the sum of the left and right electric field and integrated over the pulse duration. With our observation system (lens and CCD camera), the distance between the conjugate points of the PhC waveguide imaged by adjacent pixels is ~0.14 μm. However, for our off-chip imaging system, the actual spatial resolution is given by the diffraction limit, i.e. $\sim \lambda/(2 \times N.A.)$, resulting in ~400nm for the TH signal detected at 520nm and the objective numerical aperture (N.A.=0.65) used here. Note that this roughly corresponds to the period of the PhC, which eventually limits the attainable resolution due the Bloch mode nature of the underlying PhC waveguide mode that causes THG This translates into ~53fs (69fs) temporal resolution for the pulse duration when considering a group velocity of c/23 (c/30), as used in this particular device. Note that the temporal resolution and maximum pulse duration that can be measured with our device respectively decrease and increase with the group index of the waveguide, as illustrated on Table I. The longest pulse duration or time window mapped into space to be measured with the device can be readily extended by increasing the PhC waveguide length to the extent of dispersion remaining negligible, i.e. up to 340ps for a 2mm long PhC waveguide (in the c/30 flat-band slow light window). The shortest pulse duration available to measurement is primarily governed by the dispersion length of the PhC waveguide, i.e. here 550fs, rather than the time resolution provided by our device.

Table I. Impact of group velocity on the temporal resolution and the time window to be mapped within a 100μm or 2mm long PhC waveguide.

| Group velocity $c/n_g$ | c/8 | c/23 | c/30 |
| --- | --- | --- | --- |
| Temporal resolution: $\Delta T = \Delta Z \times n_g/c \times \sqrt{3}$ (related to a 400nm diffraction limit) | 19fs | 53fs | 69fs |
| Maximum pulse length or time window to be measured with a 100μm long device | 4.6ps | 13ps | 17ps |
| Maximum pulse length or time window to be measured with a 2mm long device | 92ps | 260ps | 340ps |

For the equivalent of "time-of-flight measurements" of Fig. 6, the tunable delay line is scanned across ~10ps, so as to change the crossing point location within the waveguide, where the right and left pulses do meet. Tracking the TH profile center location versus the delay gives a linear curve. The slope is equal to half of the group velocity, since the crossing point moves twice as slow as the pulses. This allows us to extract the average group velocity at which the pulse propagates without any assumption or fit parameter, and with a resolution limited by diffraction.

## Acknowledgements

We acknowledge the financial support of the European Union through the Marie Curie program (ALLOPTICS), as well as the Faculty of Science at the University of Sydney and the Australian Research Council (ARC).

Juntao Li is now Associated Professor at the State Key Laboratory of Optoelectronic Materials & Technologies, Sun Yat-sen University, China.

# Figure captions

Figure 1. (a) Set-up used to probe the photonic crystal waveguide in a counter-propagating configuration. (b) Schematic of the THG-based autocorrelation technique where two counter-propagating pulse replicas at the fundamental frequency interact to generate a stationary spatial profile at the third-harmonic frequency (visible) directly related to the pulse auto-correlation trace.

Figure 2. (a) Schematic of the chip including the silicon slow light photonic crystal waveguide suspended in air and connected to SU8 waveguides and inverse tapers for light coupling. (b) Group index dispersion and associated transmission of the photonic crystal waveguide. (c) Scanning Electron Image of the device.

Figure 3. Top visible images of the PhC waveguide from the CCD camera showing the TH light generated when exciting the waveguide at 1548nm from the left only (b), the right only (c), and from both sides with the off chip delay line set to 0ps (d), and 137ps (e). The dotted white line highlights the boundary of the slow light PhC waveguide where the green light is emitted. (f) Pulse intensity profile recovered from the THG profile (red) on the CCD image (e) and measured with a commercial auto-correlator (black).

Figure 4. Fourier-space images of the signal generated at the third-harmonic visible wavelength, when the counter-propagating pulses overlap off-chip (a) or into the photonic crystal waveguide (b). In the latter case, two additional lines appear in the k-space, associated with cross-THG between two (resp. one) left and one (resp. two) right photons. The conservation of photon momentum in the direction of the waveguide (quasi-phase matching along $k_x$) for the THG and cross-THG process are represented below the images, explaining the different lines observed in both cases.

Figure 5. (a) Top visible images of the waveguide obtained when launching pulses with similar durations at different wavelengths, corresponding to different group velocities, and (b) normalized TH signals across the photonic crystal waveguide length (shifted vertically for clarity). (c) displays the full-width at half maximum of the TH intensity profile fitted by a gaussian shape (red lines on (b)) versus the inverse of the group index at the different wavelengths probed.

Figure 6. (a) Top visible images of the TH signal for various relative delays between the right and left arms adjusted off-chip ($\lambda$= 1548nm). (b) Position of the TH peak intensity center along the waveguide versus the relative delay. The group velocity at which the pulse propagates can be inferred from the slope. (c) Group index measured from the TH images (green stars) overlaid on the waveguide dispersion measured by interferometric methods. The colored arrows correspond to the 4 different wavelengths probed on Figs. 5(a,b) and 6(c) while the grey areas indicate the bandwidth associated with the 2.5ps pulses.

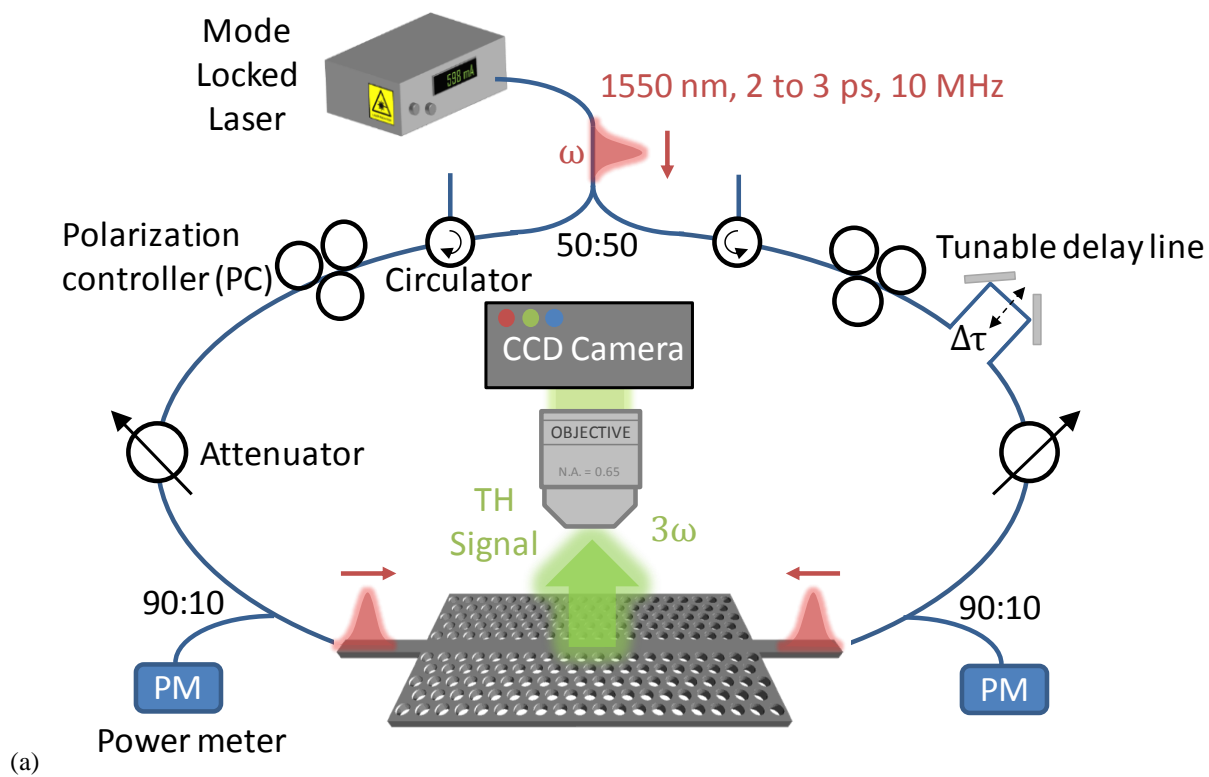

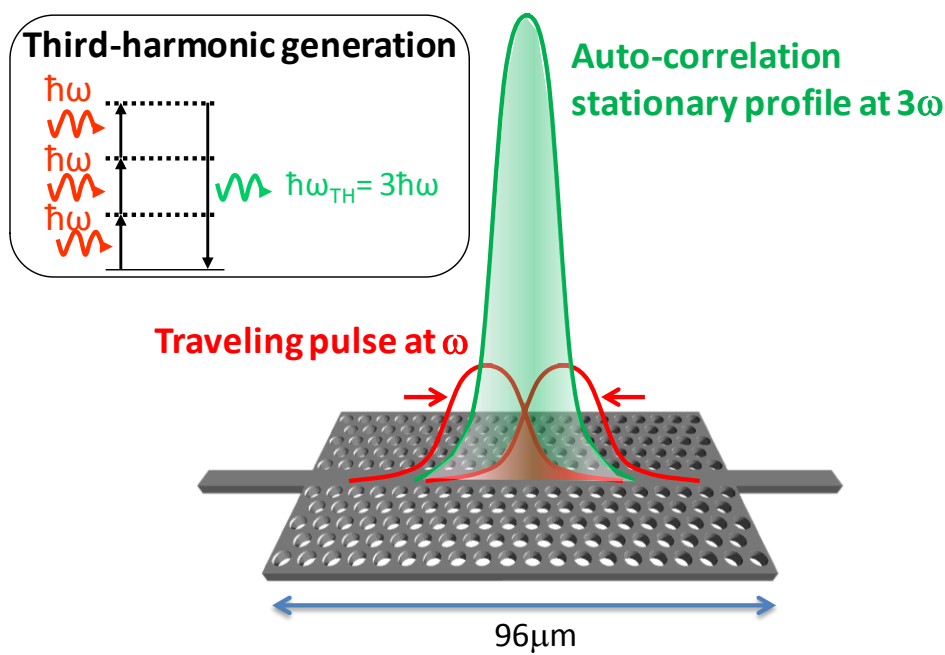

Figure 1. (a) Set-up used to probe the photonic crystal waveguide in a counter-propagating configuration. (b) Schematic of the THG-based autocorrelation technique where two counter-propagating pulse replicas at the fundamental frequency interact to generate a stationary spatial profile at the third-harmonic frequency (visible) directly related to the pulse auto-correlation trace.

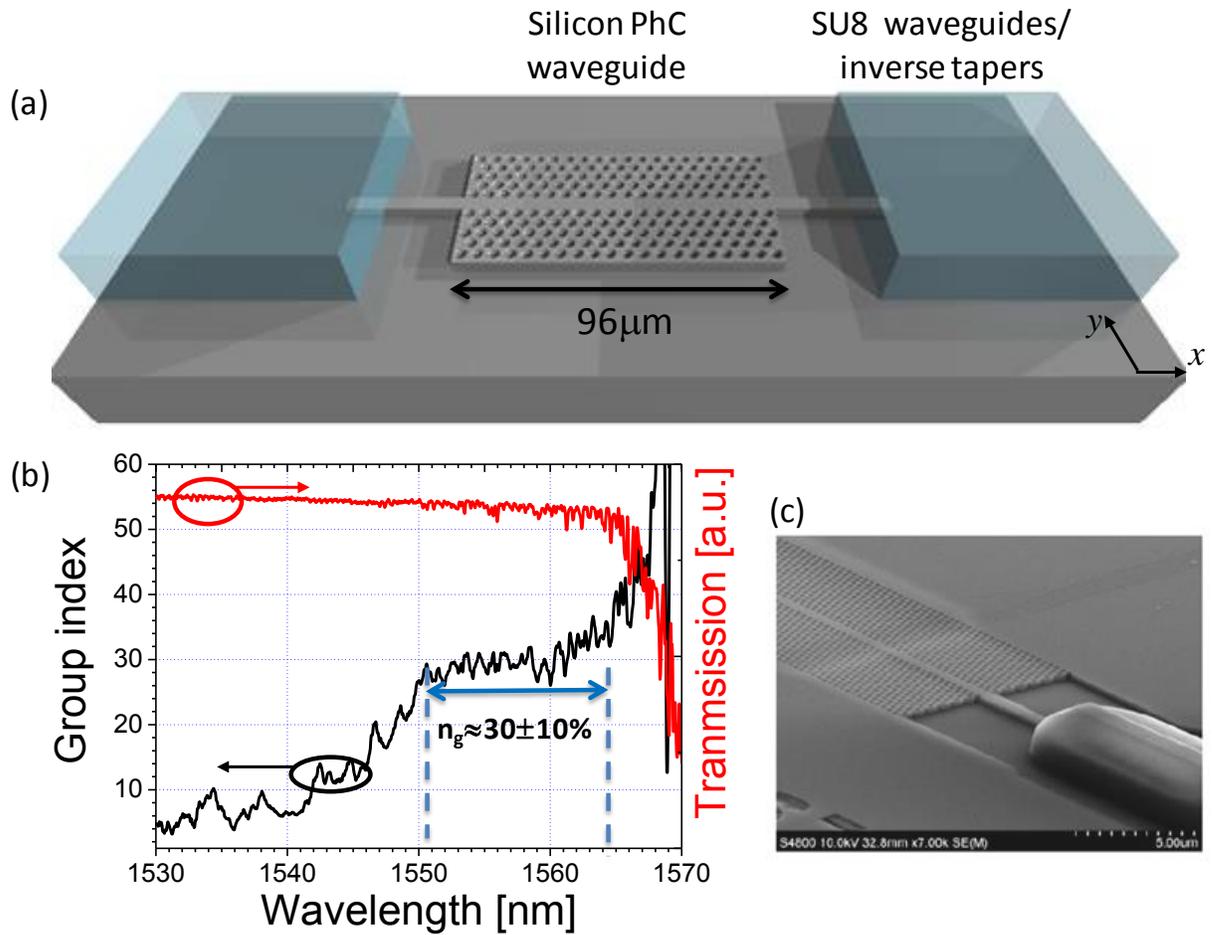

Figure 2. (a) Schematic of the chip including the silicon slow light photonic crystal waveguide suspended in air and connected to SU8 waveguides and inverse tapers for light coupling. (b) Group index dispersion and associated transmission of the photonic crystal waveguide. (c) Scanning Electron Image of the device.

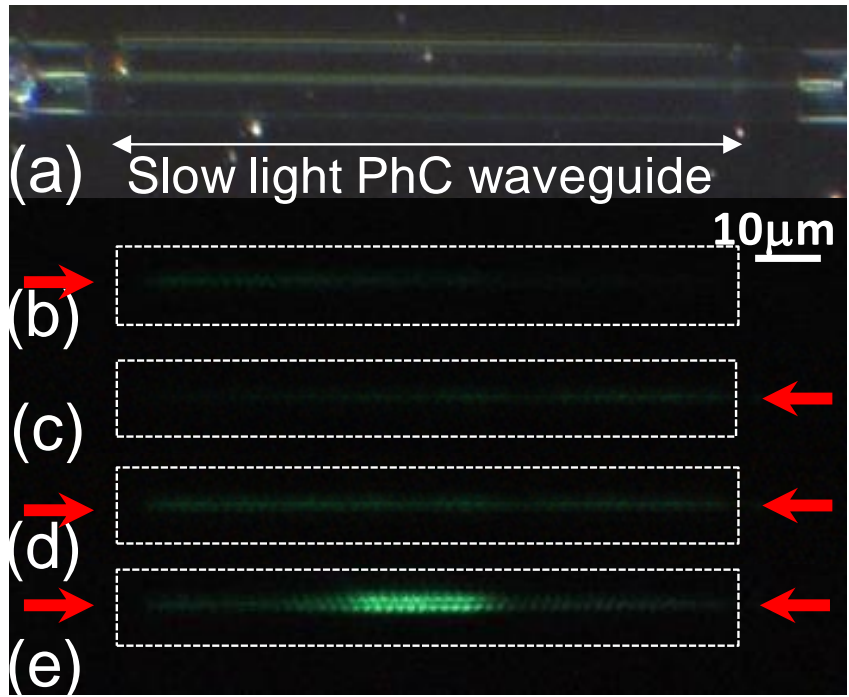
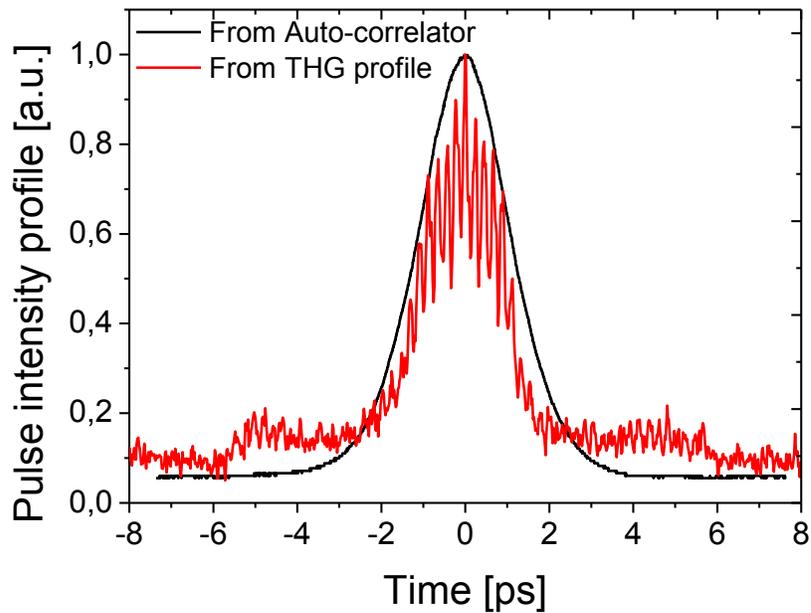

(f)

Figure 3. Top visible images of the PhC waveguide from the CCD camera showing the TH light generated when exciting the waveguide at 1548nm from the left only (b), the right only (c), and from both sides with the off chip delay line set to 0ps (d), and 137ps (e). The dotted white line highlights the boundary of the slow light PhC waveguide where the green light is emitted. (f) Pulse intensity profile recovered from the THG profile (red) on the CCD image (e) and measured with a commercial auto-correlator (black).

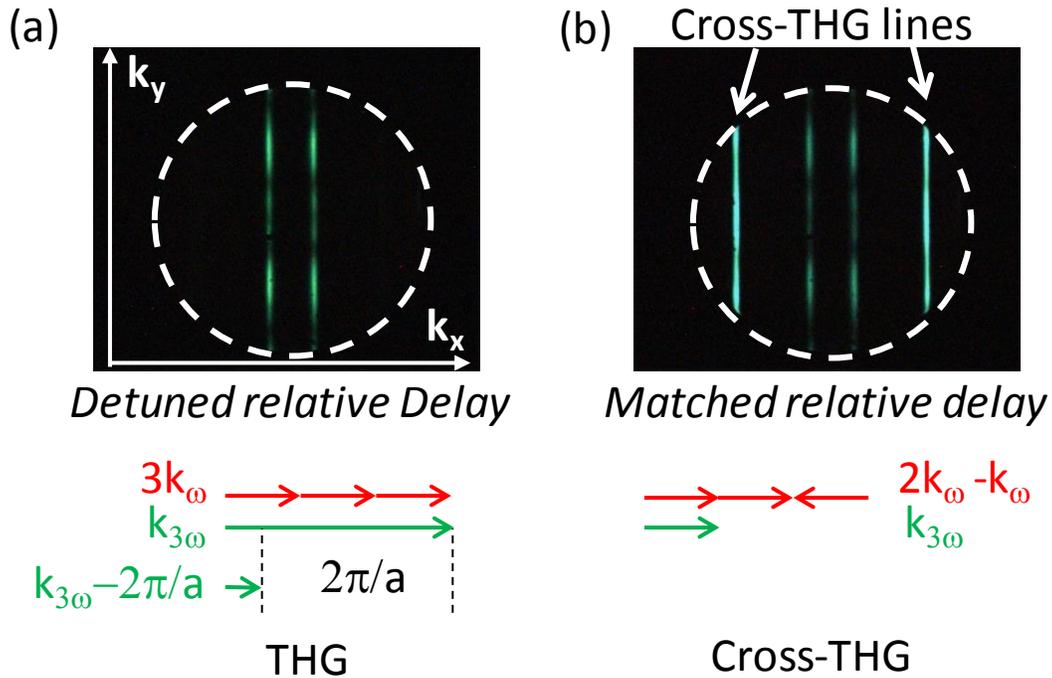

Figure 4. Fourier-space images of the signal generated at the third-harmonic visible wavelength, when the counter-propagating pulses overlap off-chip (a) or into the photonic crystal waveguide (b). In the latter case, two additional lines appear in the k-space, associated with cross-THG between two (resp. one) left and one (resp. two) right photons. The conservation of photon momentum in the direction of the waveguide (quasi-phase matching along $k_x$) for the THG and cross-THG process are represented below the images, explaining the different lines observed in both cases.

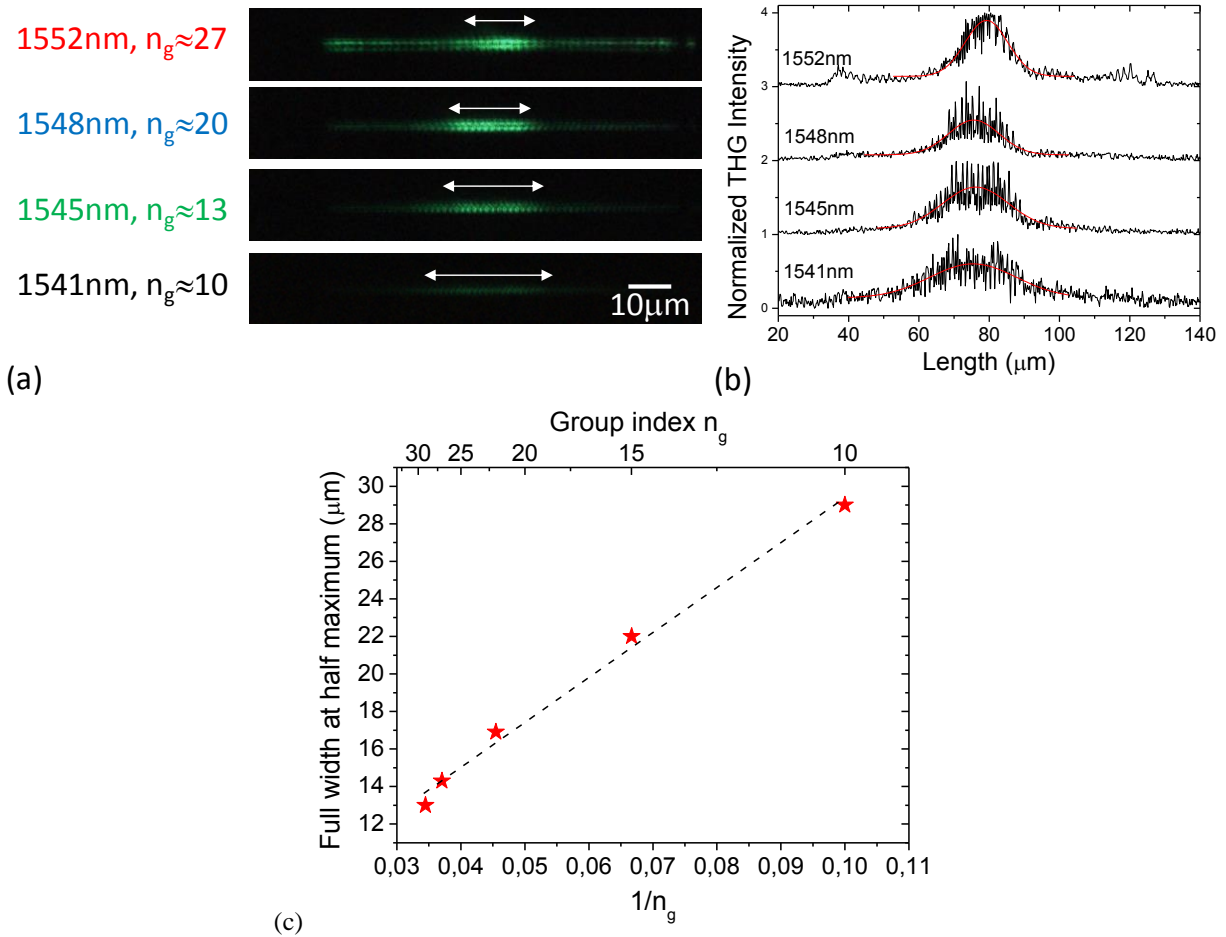

Figure 5. (a) Top visible images of the waveguide obtained when launching pulses with similar durations at different wavelengths, corresponding to different group velocities, and (b) normalized TH signals across the photonic crystal waveguide length (shifted vertically for clarity). (c) displays the full-width at half maximum of the TH intensity profile fitted by a gaussian shape (red lines on (b)) versus the inverse of the group index at the different wavelengths probed.

.

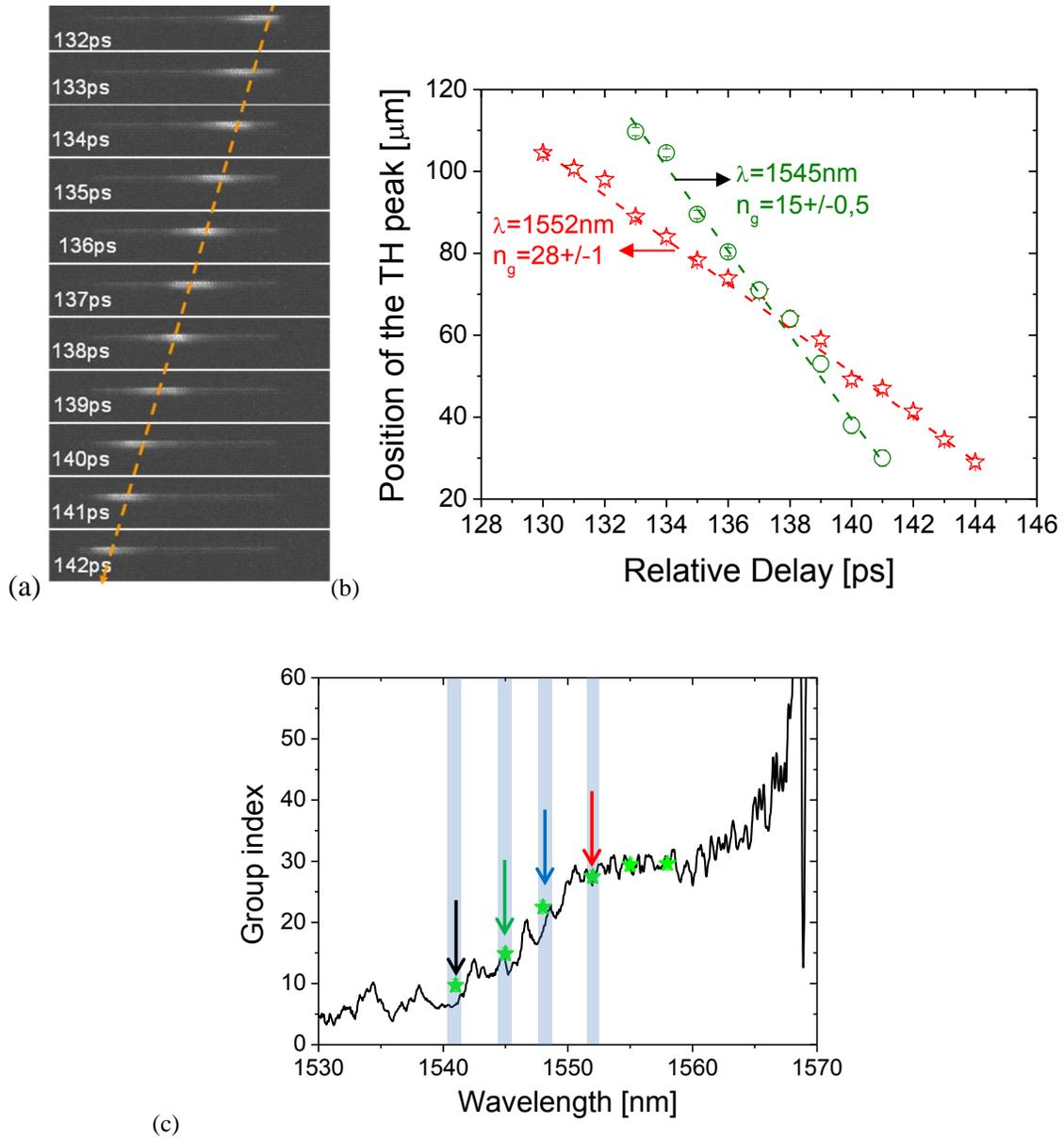

Figure 6. (a) Top visible images of the TH signal for various relative delays between the right and left arms adjusted off-chip (λ= 1548nm). (b) Position of the TH peak intensity center along the waveguide versus the relative delay. The group velocity at which the pulse propagates can be inferred from the slope. (c) Group index measured from the TH images (green stars) overlaid on the waveguide dispersion measured by interferometric methods. The colored arrows correspond to the 4 different wavelengths probed on Figs. 5(a,b) and 6(c) while the grey areas indicate the bandwidth associated with the 2.5ps pulses.